\documentclass{PoS}

\title{AMS tracking in-orbit performance}

\ShortTitle{AMS tracking in-orbit performance}

\author{\speaker{Martin Pohl}%
        \thanks{On behalf of the AMS-02 Collaboration}\\
       DPNC, University of Geneva\\
       E-mail: \email{martin.pohl@cern.ch}}

\abstract{AMS-02 is a high precision magnetic spectrometer for cosmic rays in the GeV to TeV energy range. Its tracker consists of nine layers of double-sided silicon microstrip sensors. They are used to locate the trajectories of cosmic rays in the 0.14 T field of a cylindrical magnet, thus measuring their rigidity $p/Z$ and charge sign. In addition, they deliver a high resolution measurement of the absolute charge $|Z|$. The detector has been designed to operate in space with a position resolution of about 10 $\mu$m for each hit and charge identification capabilities up to $Z=26$. In this talk I describe the performance in orbit of this detector component and its impact on the overall performance of the spectrometer.}

\FullConference{24th International Workshop on Vertex Detectors\\
		 1-5 June 2015\\
		 Santa Fe, New Mexico, USA}

\begin{document}

\section{Introduction}
Space astroparticle missions analyse cosmic rays and high energy photons. 
Cosmic ray composition and spectra are measured in near Earth orbit, typically in the 100 MeV to multi-TeV energy range. The momentum vector, including arrival direction and time, as well as mass and charge of the particle completely characterize its properties. The arrival direction is needed to distinguish primary cosmic rays from those trapped in the Earth magnetic field, as well as to search for anisotropies, even though charged particules do not point back to their origin but are scrambled by magnetic fields at all scales. 
Photons, on the other hand, do point back to their point-like, extended or diffuse sources. Astroparticle physics is interested in photons from non-thermal processes, typically in the X-ray to TeV energy range. In addition to the energy, arrival direction and time, the degree and direction of polarisation provides valuable information. The two research fields are linked by the fact that high energy photons come from sources where cosmic rays are generated and accelerated. An example of a high energy photon detector is the Large Area Telescope LAT~\cite{Fermi} of the Fermi satellite, in orbit since June 11, 2008. The Alpha Magnetic Spectrometer AMS-02, in orbit since May 16, 2011, sets a new scale in the detection of cosmic rays. 

Cosmic ray composition, spectra per species and anisotropies near Earth carry information about the production, acceleration and transport of cosmic rays in the galaxy. Among the most interesting research subjects is the search for non-astronomical sources and unusual components of cosmic rays. Experimental conditions are characterized by  low rates, and thus low occupancy in the detector. However, other requirements are much the same as those found in collider experiments. The layered structure and redundancy of space astroparticle detectors is thus similar to those of modern collider detectors. 

Cosmic radiation contains about 87\% protons, 9\% He, few percent heavy ions, and even fewer e$^\pm$ and photons. Fluxes fall like power laws, typically by three orders of magnitude per decade in energy. Because of this rapidly falling flux, the acceptance in terms of m$^2$sr determines the energy reach of a detector, thus size matters. Detectors for high energy photons must typically exceed the acceptance of cosmic ray detectors. However, obvious limitations of size, weight and power consumption are imposed by experimentation in space. In addition, the inaccessibility of the hardware in space requires a reliability rarely met by terrestrial instruments~\cite{TIPP}. 

\section{Launch and in-orbit requirements}\label{sec:requirements}
The mechanical requirements on space hardware are dominated by the harsh launch process. High levels of acoustic noise, static and vibrational loads are encountered. Moreover, pyro-shocks during the separation of rocket components can reach levels of several thousand $g$ and high frequencies in unmanned space flights. It is thus mandatory to qualify equipment for both longitudinal and lateral vibrations and shocks, and to determine resonant frequencies. Qualification levels usually include a large margin of safety compared to actual launch conditions. Other requirements concern interactions with the space craft and other payloads, like cleanliness, particle and chemical contaminations and electromagnetic compatibility. Details of the qualification requirements are given in the documentation of each launch vehicle, for popular ones see~\cite{ref_manuals}.  

The in-orbit environment is unusually hostile for sensitive particle detectors. Micrometeorites and orbiting debris may impact at the speed and mass of a gun projectile. Atomic oxygen may corrode surfaces. Charging by plasma is an issue, especially when transiting the Earth radiation belts. Irradiation~\cite{ref_Bourdarie} by protons and electrons trapped in the Earth magnetosphere, by solar particles and cosmic rays themselves may cause latch-up or single event effects. Usage of modern electronic components or sensors thus often requires qualification for the expected environment. 

The thermal environment in space often presents the most demanding challenge, since it depends on many variables. Orbital properties determine the periodic day/night transition. The angle $\beta$ between the orbital plane and the solar vector determines illumination conditions. Radiators and solar panels of the space craft may shadow the payload. The payload attitude may suddenly change by space craft maneuvers. All of this introduces thermal changes with very different time scales, from minutes to months. The result may be thermo-mechanical deformations requiring dynamic alignment. Properties of electronic components like noise levels and gains may shift as a function of temperature, requiring frequent calibration. And ultimately, damaging effects may occur if excursions outside the operational or even survival temperature range of the equipment are encountered. Thermal control can in the best case be implemented passively by heat conduction towards radiators. Active thermal control can require pumps and pressurized systems which are more difficult to design for space applications. Payload operations on ground may not always be able to control the thermal environment at any given time. It is thus wise to implement automatic safeguard procedures.

\section{The Alpha Magnetic Spectrometer on the ISS}
\begin{figure}[!ht]
\begin{center}
\includegraphics[width=0.4\columnwidth]{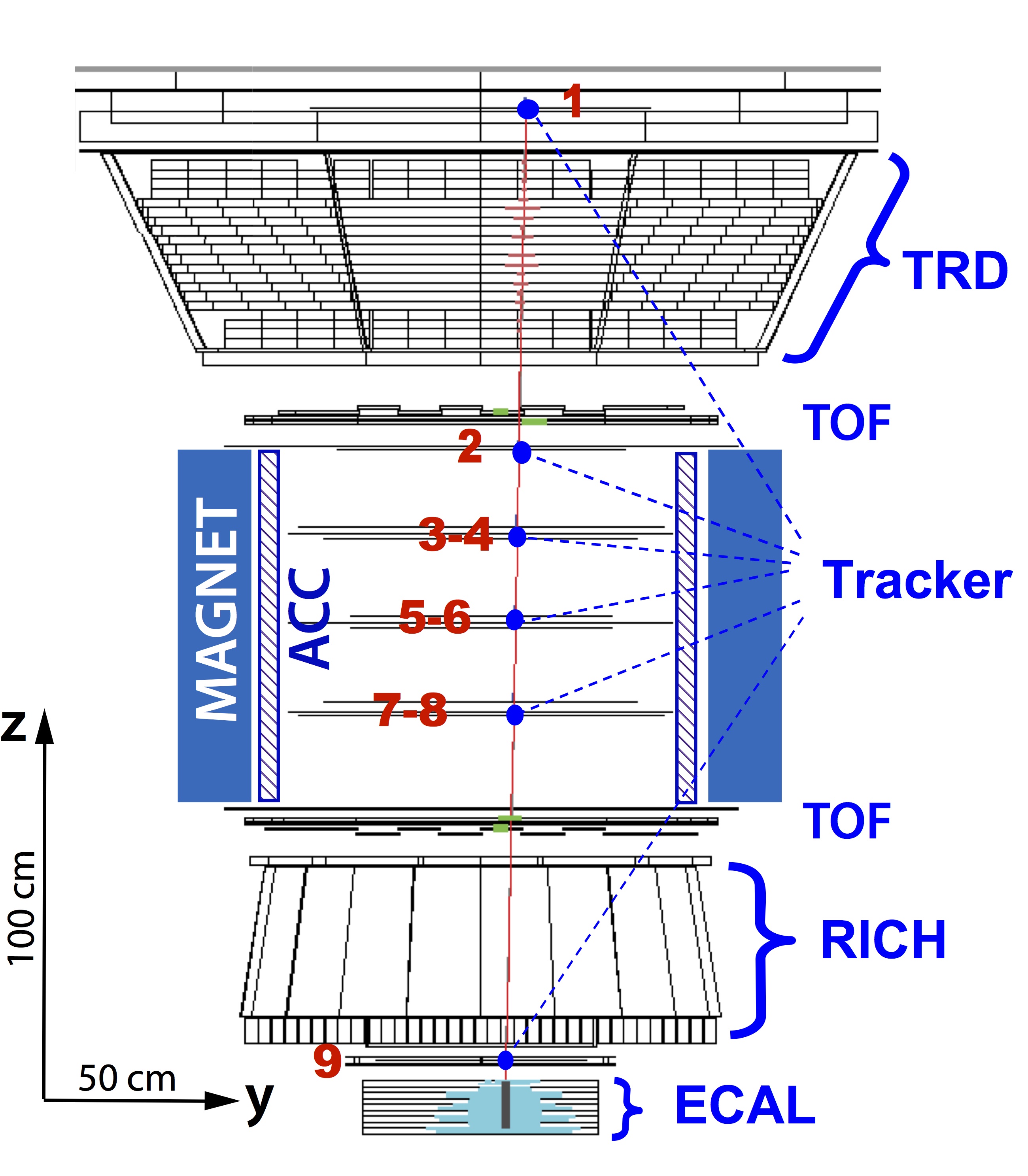}
\caption{A 369 GeV positron event as measured by the AMS detector on the ISS in the bending plane. Tracker layers 1 to 9 measure the particle charge, sign, and momentum. The TRD identifies the particle as e$^\pm$. The TOF measures the absolute charge value to be 1 and ensures that the particle is downward-going. The RICH independently measures the charge and velocity. The ECAL measures the 3D shower profile, independently identifies the particle as an e$^\pm$ and measures its energy. A positron is identified by 1) positive rigidity in the tracker, 2) an e$^\pm$ signal in the TRD, 3) an e$^\pm$ signal in the ECAL and 4) the matching of the ECAL shower energy and axis with the momentum measured with the tracker and magnet.}
\label{fig:positron_event}
\end{center}
\end{figure}

The layout of the AMS-02 detector~\cite{AMS_detector} is shown in Fig.~\ref{fig:positron_event}. It consists of 9 layers of precision silicon strip detectors with two outer layers, 1 and 9, and the inner tracker, layers 2 to 8; a transition radiation detector, TRD; four planes of time of flight counters, TOF; a permanent magnet with a central field strength of 0.15 T; an array of anti-coincidence counters, ACC, inside the magnet bore; a ring imaging Cherenkov detector, RICH; and an electromagnetic calorimeter, ECAL. The figure also shows a high energy positron of 369 GeV recorded by AMS. AMS operates without interruption on the ISS and is monitored continuously from the ground. The maximum detectable rigidity $p/Z$ over tracker layers 1 through 9, with a lever arm of 3 m, is about 2 TV. Detector performance is steady over time.

\begin{figure}[!ht]
\begin{center}
\includegraphics[width=0.7\columnwidth]{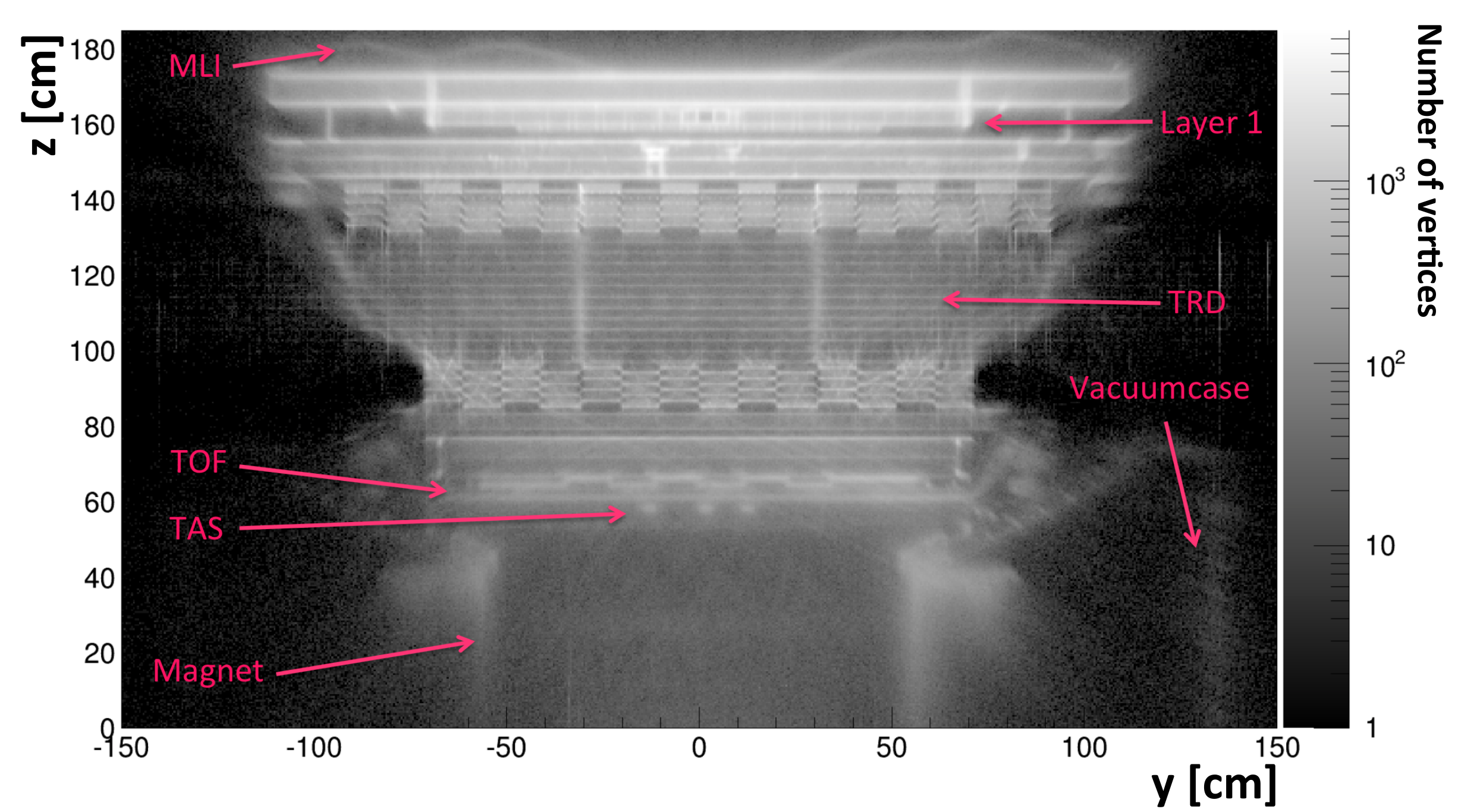}
\caption{Distribution of vertices reconstructed from multiprong events in AMS. The distribution of material is seen in great detail, including fine structures like the plastic tubes of the TRD. AMS Collaboration courtesy A. Bachlechner, RWTH Aachen University.}
\label{fig:xray}
\end{center}
\end{figure}

Recent physics highlights from AMS include the measurement of the electron and positron spectra, their sum and their ratio~\cite{AMS-epm} as well as those of primary protons~\cite{AMS_p}. Many other results are in preparation concerning light~\cite{AMS_he} and heavy nuclei as well as antiprotons~\cite{AMS_ICRC_2015}. As an example of the excellent performance of the tracking device, Fig.~\ref{fig:xray} shows the position of vertices reconstructed from rare multi prong events, which clearly show the distribution of material over the volume of the detector. This information is used to refine the description of the detector geometry and materials in its GEANT4 simulation~\cite{GEANT4}.

\section{The AMS tracking detector}

The AMS-02 silicon strip tracking detector inherits from proven technology used for the LEP~\cite{L3_SMD} and LHC~\cite{ATLAS_SCT}  collider experiments to construct micro vertex detectors, but also from other astroparticle missions~\cite{Fermi_LAT}.  This technology continues to be used in state-of-the-art astroparticle instruments~\cite{DAMPE}.
In contrast to others, the AMS-02 tracker is constructed from sensors with double-sided readout. It consists of 9 layers allowing for the measurement of 9 3D hits along an incoming particle's trajectory with a single point precision of 10 microns. Each hit records bending and non-bending coordinates, as well as two measurements of the specific energy loss $dE/dx$. 

Each layer is made up of multiple ladders with a total number of 192 ladders. The ladders consist of 10 to 15 sensors, totaling 2284 double sided silicon sensors, each $72 \times 41$ mm$^2$ large and 300 $\mu$m thick. 
In order to reduce the amount of channels,  strips measuring the bending coordinate with a readout pitch of 110 $\mu$m are daisy-chained along the ladder length and readout via 10 VAhdr9a ASICs of the Viking family. Orthogonal strips with a readout pitch of 208 $\mu$m are multiplexed through a kapton cable that connects them to the front-end board holding 6 VAhdr9a chips. 
In total, the tracker has about 200.000 readout channels. The ASIC has non-linear response characteristics to increase the dynamic range of the pulse height measurement, with different gain for bending and non-bending direction. Excellent charge resolution is thus obtained from hydrogen up to iron.  An early zero suppression is performed by the Tracker Data Reduction boards. The readout electronics were designed to have a small power consumption (0.7 mW per channel) and low noise. More details on design and production of the tracking device can be found in \cite{Azzarello}. 

\begin{figure}[!ht]
\begin{center}
\includegraphics[width=0.6\columnwidth]{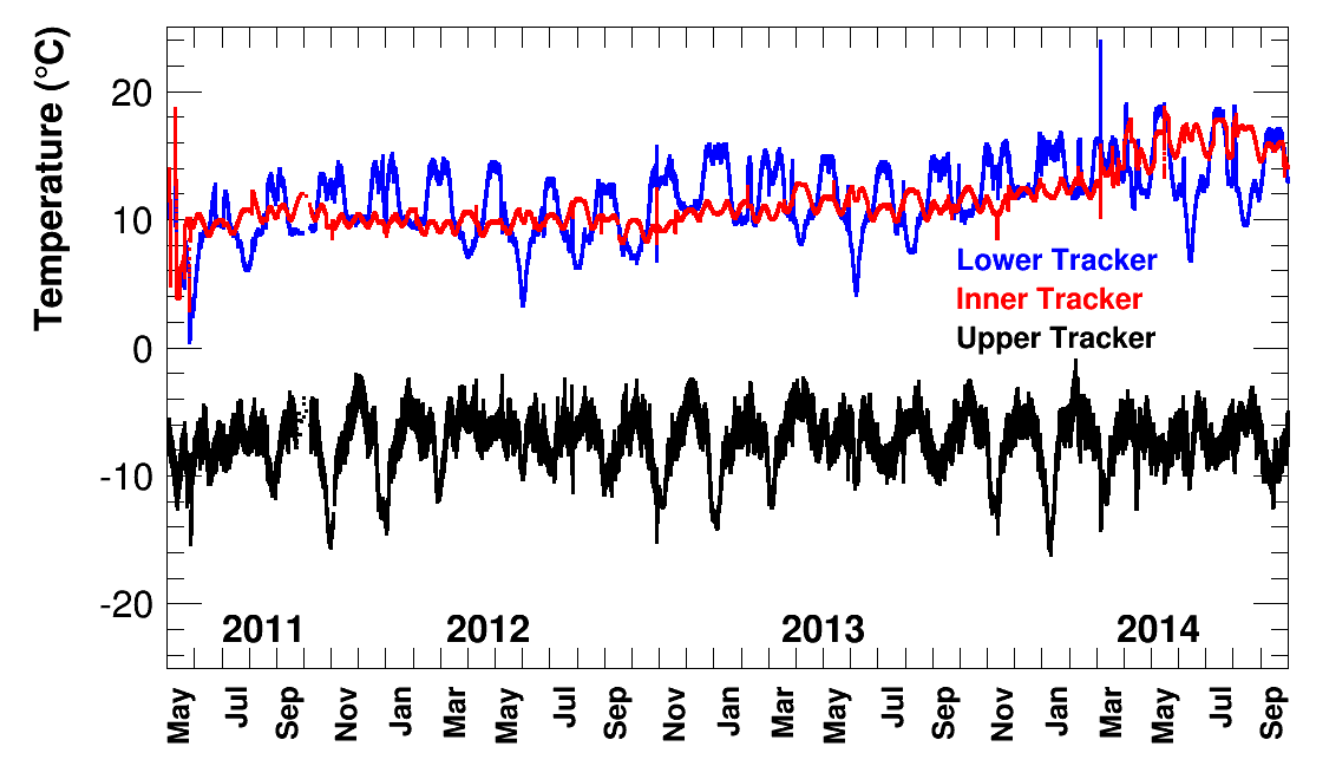}
\caption{Temperatures measured on layer 1 (black), layers  2 through 8 (red) and 9 (blue). The inner tracker temperature is actively controlled and stable except for very long term variations. Layer 1 is passively cooled and shows larger short term variations. Layer 9 is in radiative and conductive thermal contact with RICH and ECAL. }
\label{fig:temp}
\end{center}
\end{figure}

The thermal environment on the ISS is demanding. The thermal variables related to orbit and space craft, mentioned in Section~\ref{sec:requirements}, lead to temperature gradients up to  $\pm$20 degrees on time scales ranging from months to minutes. In addition, the total power consumption of the tracker is about 200 W. The temperature inside the magnet bore is thus actively controlled by the Tracker Thermal Control System (TTCS)~\cite{ref_Vanes}.  The readout electronics of layers 2 to 9 are connected through thermal bars to two cooling loops filled with high-pressure liquid CO$_2$. The CO$_2$ absorbs the heat by evaporating a small portion of the liquid. Radiators are connected to the other side of the cooling loops, where the CO$_2$ gives off the absorbed heat and returns to its liquid phase. The layers inside the magnet bore benefitting from this active temperature control show variations below about 1 degree, as shown in Fig.~\ref{fig:temp}. The outermost layer 1 only receives passive temperature control, leading to larger excursions. The lowest layer 9 is in radiative and conductive thermal contact with RICH and ECAL. 

\begin{figure}[!ht]
\begin{center}
\includegraphics[width=0.7\columnwidth]{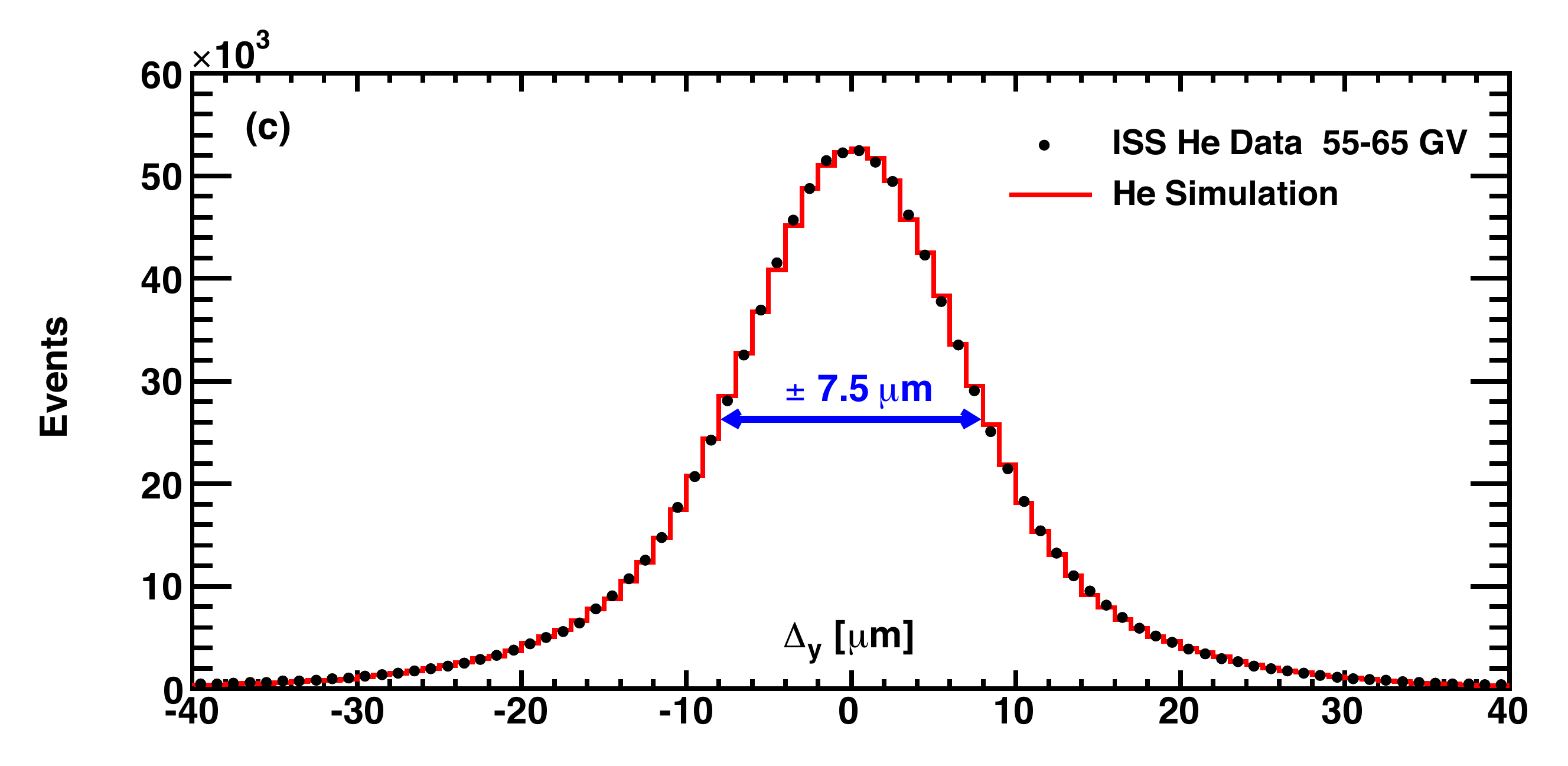}
\caption{Distribution of AMS track fitting residuals for particles with charge 2 and rigidity between 55 and 65 GV. Black dots show the data distribution, the red histogram is the result of Monte Carlo simulation. The half width at half maximum is 7.5 $\mu$m.}
\label{fig:res}
\end{center}
\end{figure}

The variation of the mean temperature and gradients across the mechanical structure lead to shifts of several hundred $\mu$m which vary on the above mentioned time scales. We have thus developed two statistically independent alignment procedures to correct the effect of these thermal movements dynamically using in-flight data, mainly from protons and helium. A first one uses a sliding time window to estimate the required correction. The second method folds several consecutive orbits to build a model of the layer movements due to the thermal variation and to compute the correction. Both methods have similar performance and are almost statistically independent, allowing their combination. The resulting tracker resolution is well understood down to its small tails, as shown in Fig.~\ref{fig:res}.

\section{Rigidity and charge measurement}

\begin{figure}[!ht]
a) \hspace*{68mm} b)

\vspace*{-10mm}
\begin{center}
\raisebox{5mm}{\includegraphics[width=0.44\columnwidth]{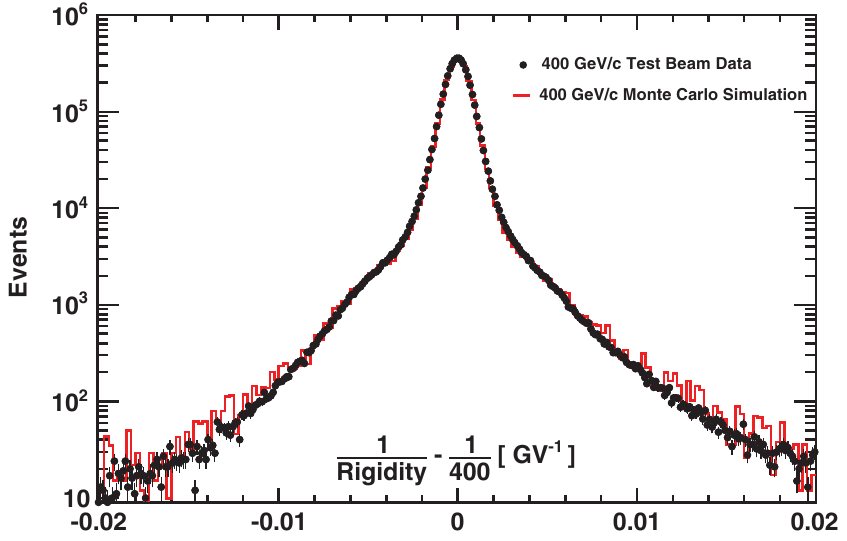} } \hspace*{7mm}
\includegraphics[width=0.48\columnwidth]{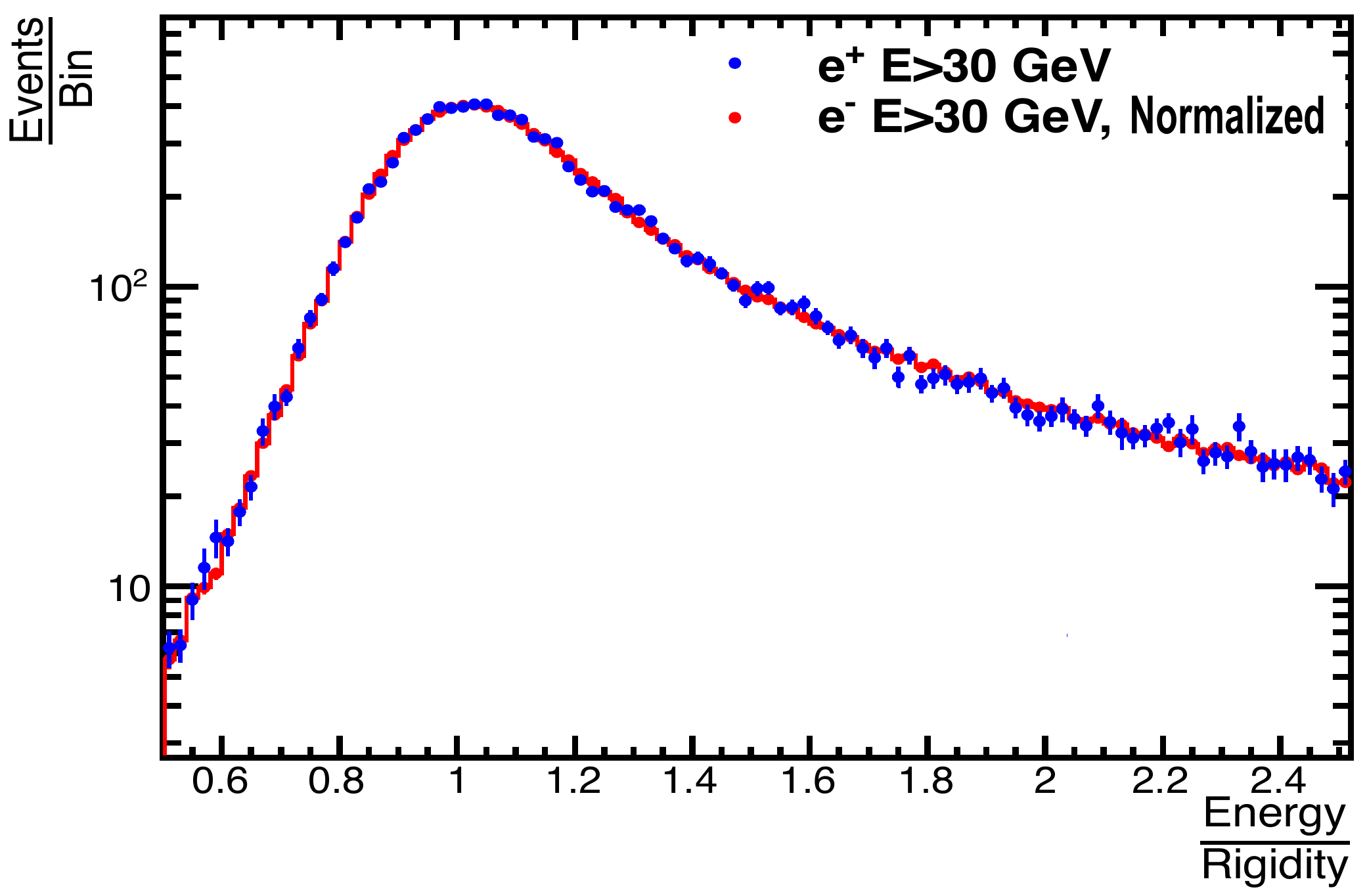} 
\caption{a) Rigidity resolution measured in a CERN SPS beam of 400 GeV protons before launch. b) Ratio between energy and rigidity for high energy electrons (red histogram) and positrons (blue dots) measured  on the ISS.  }
\label{fig:TB_res}
\end{center}
\end{figure}

Since cosmic ray fluxes are all falling rapidly with rigidity, it is important to understand the rigidity resolution of the spectrometer, which is used in the unfolding procedure for all spectra. The resolution function is taken from Monte Carlo simulation extensively cross-checked using pre-flight and flight data. As an example, Fig.~\ref{fig:TB_res}a)  shows the rigidity resolution for 400 GeV protons measured in a CERN SPS test beam before flight. The Gaussian core and the tails of the distribution are well reproduced by the simulation. Since the in-flight alignment and calibration is mainly based on positive curvature tracks, it is important to cross-check that no rigidity bias is introduced by the procedure. High statistics pre-flight data show an upper limit on the rigidity bias of $\Delta(1/R) \leq 1/300$ TV$^{-1}$. Fig.~\ref{fig:TB_res}b) shows the in-flight cross-check using electrons and positrons of greater than 30 GeV energy and measuring the ratio $E/R$ of energy measured in the electromagnetic calorimeter over rigidity measured by the spectrometer. We find an upper limit for a residual rigidity bias of $\Delta(1/R) \leq 1/26$ TV$^{-1}$, an order of magnitude smaller than the maximum determined rigidity\footnote{The maximum determined rigidity MDR is defined as the rigidity at which the error reaches 100\%.}  of 2 TV for protons and a few TV for heavier elements. The accuracy of this cross-check is limited by the e$^\pm$ statistics. For protons above 1 TV, the systematic flux error coming from the rigidity scale is about 4\%, the unfolding error due to uncertainties in the resolution function contributes about 3\% to the total systematics.

\begin{figure}[!ht]
\begin{center}
\includegraphics[width=0.6\columnwidth]{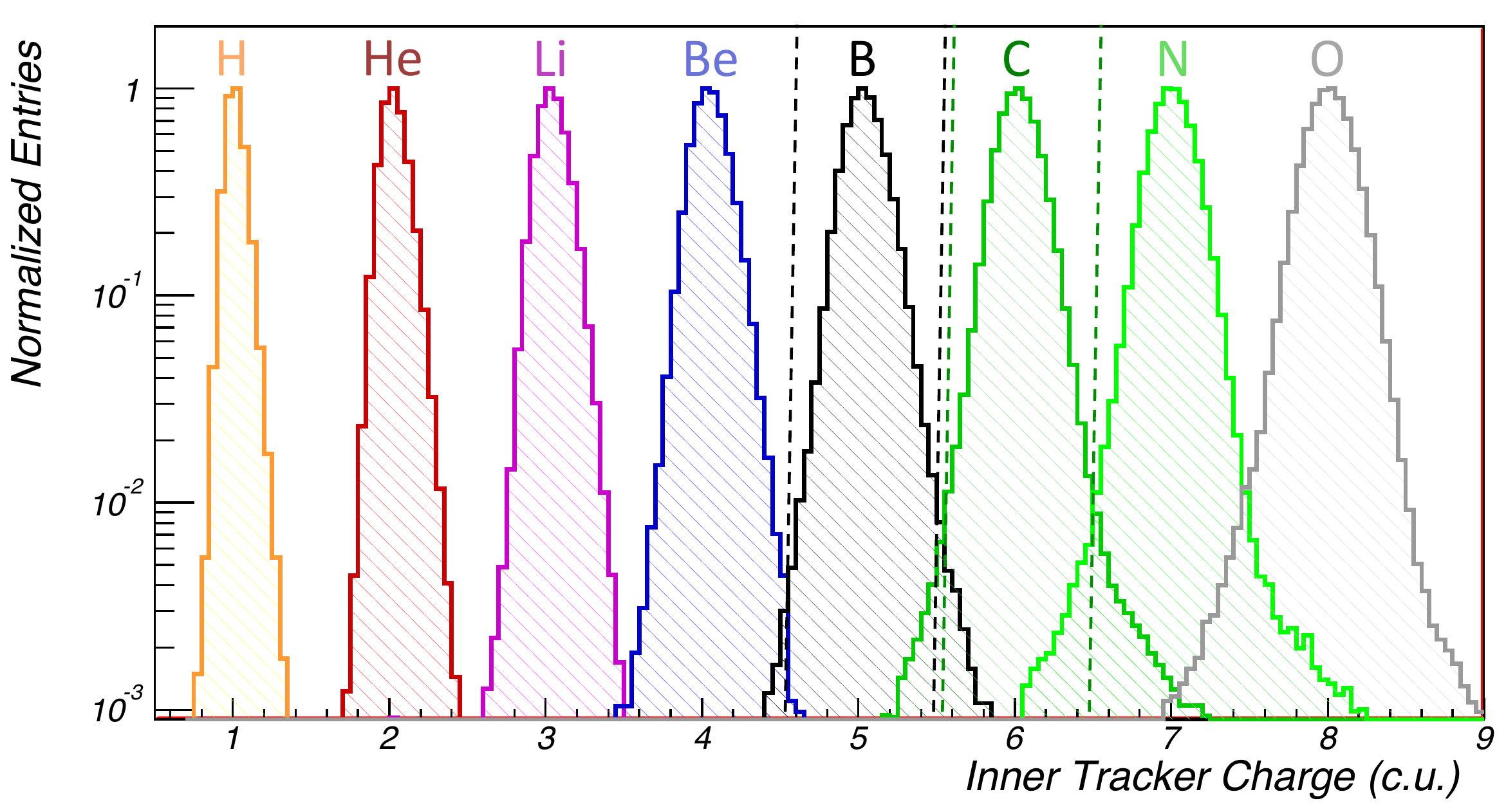}
\caption{Normalized distributions of nuclear charge measured by the inner tracker planes, for unbiased samples of nuclei selected using the rest of the detector.}
\label{fig:Z_inner}
\end{center}
\end{figure}

\begin{figure}[!ht]
\begin{center}
\includegraphics[width=0.65\columnwidth]{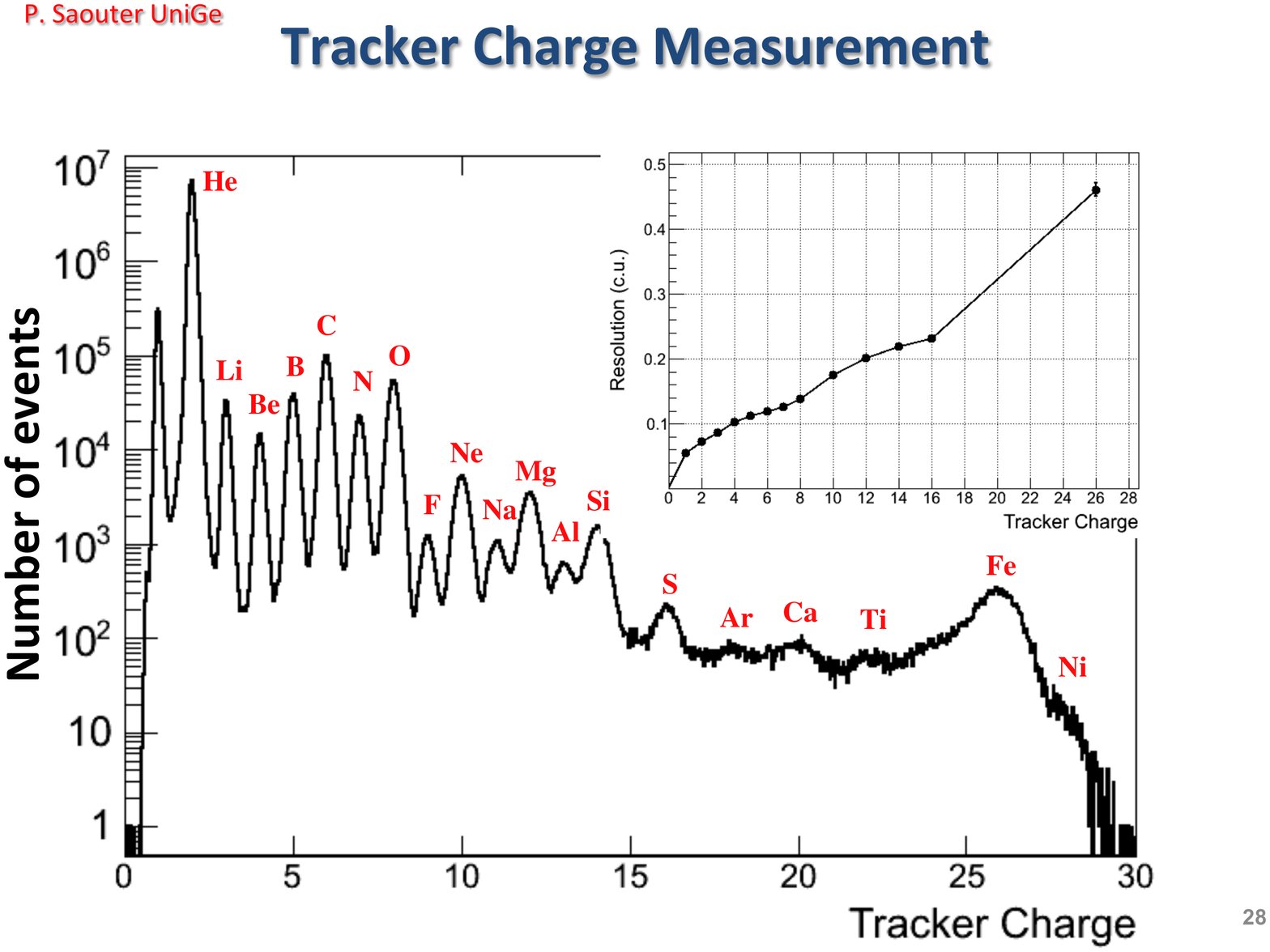}
\caption{Distribution of nuclear charge measured by the whole tracker, for a sample with strong suppression of $Z=1$ particles. The insert shows the resolution of the charge measurement in charge units.}
\label{fig:Z_total}
\end{center}
\end{figure}

Redundancy is also key to the high fidelity measurement of nuclear charge by $dE/dx$. The specific energy loss is measured no less than 13 times along the particle trajectory. Typical charge resolutions for Carbon nuclei after calibration are 0.3 charge units per tracker layer, 0.33 in the TRD, 0.16 in the upper and lower time-of-flight system, 0.32 in the RICH counter. 
Each tracker hit delivers two measurements of specific energy loss, with different gain and saturation properties on the two readout sides. The charge calibration procedure~\cite{Saouter} corrects for path length differences, the front-end gains, charge losses depending on impact position and angle as well as the dependence of $dE/dx$ on momentum. It uses unbiased samples of nuclei selected with the rest of the detector. In this way, a charge measure is determined for every hit which can be used independent of track geometry and rigidity. Fig.~\ref{fig:Z_inner} shows the truncated mean of this quantity for hits in the inner tracker, layers 2 to 8. The misidentification from neighboring charges is less than $10^{-3}$ at a charge identification efficiency of greater than 98\%. Combining with hits in the outermost layers 1 and 9, one obtains the raw charge distribution for elements from helium to the iron group shown in Fig.~\ref{fig:Z_total}. The measurements of the spectra of heavy elements in cosmic rays are in progress and will be published soon. 

\section{Outlook}
\begin{figure}
\begin{center}
\includegraphics[width=0.5\textwidth]{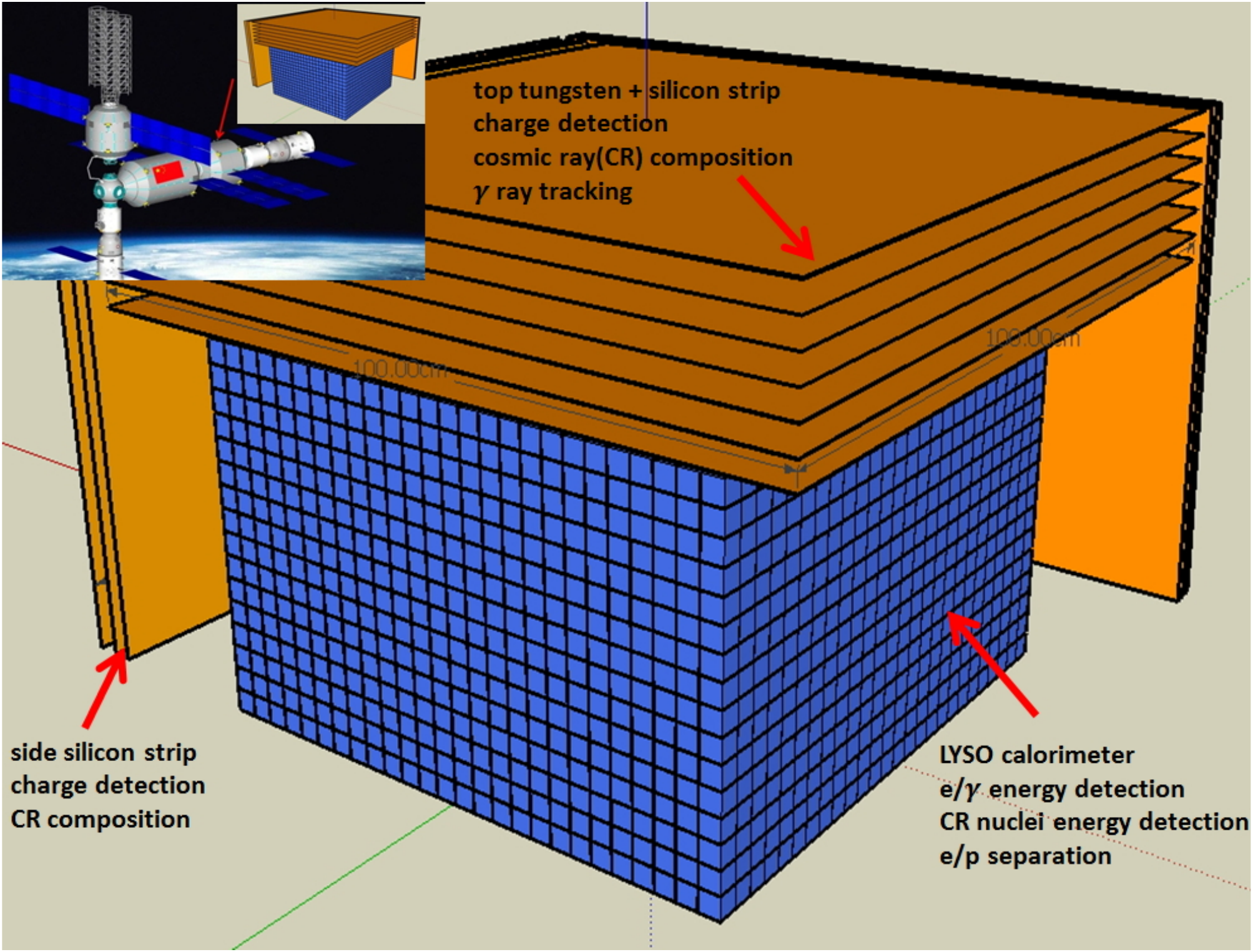}
\end{center}
\caption{Conceptual design of the layout and accommodation of HERD. The heart of the detector is a 3D calorimeter made of 3$\times$3$\times$3 cm$^3$ LYSO cubes read out by fibers via an image intensifier. The set-up is surrounded on 5 sides by silicon strip charged particle detectors, layered on top with tungsten plates for early photon conversion.}
\label{fig:herd} 
\end{figure}

The next generation of astroparticle experiments in space will not only be bigger and better but also use new ideas. The missions are becoming even more like general purpose particle physics experiments, in that they try to cover cosmic ray physics, dark matter search and gamma ray astronomy in the same mission. They are thus striving to measure photons, electrons, protons and heavy ions all with the same payload, usually with calorimetric detectors. Several missions are approved and will be installed soon~\cite{CREAM, CALET, DAMPE}, more are in the planning phase~\cite{GAMMA-400}. The next big leap forward might easily be the High Energy Radiation Detector HERD~\cite{HERD}. This mission, planned for the Chinese Space Station in the 2020s will measure electrons and photons from 100 GeV to 10 TeV and extend cosmic ray flux measurements from 300 GeV to the ``knee'' at PeV energies. It will thus finally cover the energy where the general cosmic ray flux changes slope with a direct composition measurement. It will also help gamma ray astronomy by monitoring GRB and other transients. It may accommodate an extension to lower energies, towards MeV to GeV photons with an instrument like PANGU~\cite{PANGU}. The heart of the detector is a large 3D calorimeter surrounded by tracking devices with an acceptance of several m$^2$sr. The conceptual design of HERD is shown in Fig.~\ref{fig:herd}. The advantage of 3D calorimetry is that it avoids a layered structure and can thus be sensitive in a field-of-view of almost $2\pi$. 

\section{Conclusions}
With the success of the FERMI and AMS missions, astroparticle physics in space has entered into a new era of precision measurements. Energies approach TeV for electrons and photons and multi-TeV for ions. This opens the way to important synergies with ground-based astroparticle experiments.Three major missions~\cite{CREAM, CALET, DAMPE} will go into operation soon, with the aim to improve energy resolution and acceptance in the TeV regime. The High Energy Radiation Detector~\cite{HERD} with its novel concept of 3D calorimetry may well be the next step forward, featuring large acceptance and good energy resolution up to the PeV regime, offering high precision measurements of electrons, photons and ions at the same time. Additional lever arm will come from new efforts to measure photons in the difficult keV to MeV range, to observe X-ray transients, polarization and spectroscopy relevant for cosmic ray acceleration. These use new enabling technologies for the detection of X-rays and scintillation high detection~\cite{TIPP}. There is thus an exciting program of multi-messenger, multi-wavelength astroparticle physics research ahead of us in the next 10 to 15 years.  

\section*{Acknowledgements}
I would like to thank the organizers of Vertex 2015 for inviting me to this inspiring conference at an exciting venue. The University of Geneva AMS group gratefully acknowledges many years of support from the Swiss National Science Foundation as well as cantonal and federal authorities.


\begin{thebibliography}{99}
 \bibitem{Fermi} E. Charles, {\em Lessons from the Fermi LAT experience for high precision trackers for future space missions}, these proceedings. 
 \bibitem{TIPP} See e.g. M. Pohl, {\em Particle detection technology for space-borne astroparticle experiments}, PoS TIPP2014 (2015) 013. 
 \bibitem{ref_manuals} Arianespace Soyuz Users Manual (2012), {\footnotesize \verb+www.arianespace.com/launch-services-soyuz/Soyuz-User's-Manual.asp+}; \\
Arianespace Ariane 5 User's Manual (2011), {\footnotesize \verb+www.arianespace.com/launch-services-ariane5/Ariane-5-User's-Manual.asp+};
Arianespace Vega Users Manual (2006), {\footnotesize \verb+www.arianespace.com/launch-services-vega/Vega-user's-manual.asp+} ;
Falcon 9 Launch Vehicle Payload User's Guide (2009), {\footnotesize \verb+decadal.gsfc.nasa.gov/pace-201206mdl/Launch Vehicle Information/+}\\{\footnotesize \verb+Falcon9UsersGuide_2009.pdf+}.
\bibitem{ref_Bourdarie} S. Bourdarie and M. Xapsos, IEEE Trans. Nucl. Science 55 (2008) 1810.
\bibitem{ref_Vanes} For an innovative example see e.g. J. van Es and P. Dieleman, International Astronautical Federation Congress 2009, IAC-09.C2.7.1, {\footnotesize \verb+www.iafastro.net/iac/archive/browse/IAC-09/C2/7/3011/+}.
\bibitem{AMS_detector}
S. Rosier-Lees, in Proceedings of Astroparticle Physics TEVPA/IDM, Amsterdam 2014 (to be published); S.C.C. Ting, Nucl. Phys. B, Proc. Suppl. 243-244, 12 (2013);  A. Kounine, Int. J. Mod. Phys. E 21 1230005 (2012);  B. Bertucci, Proc. Sci., EPS-HEP (2011) 67; M. Incagli, AIP Conf. Proc. 1223 (2010) 43; R. Battiston, Nucl. Instrum. Meth. A 588 (2008) 227.
\bibitem{AMS-epm} AMS-02 Collaboration, M. Aguilar et al., Phys. Rev. Lett. 113 (2014) 121101, Phys. Rev. Lett. 113 (2014) 121102, Phys. Rev. Lett. 113 (2014) 221102.
\bibitem{AMS_p} AMS-02 Ciollaboration, M. Aguilar et al., Phys. Rev. Lett. 114 (2015) 171103
\bibitem{AMS_he} AMS-02 Ciollaboration, M. Aguilar et al., {\em Precision Measurement of the Helium Flux in Primary Cosmic Rays from Rigidities of 1.9 GV to 3 TV with the Alpha Magnetic Spectrometer on the International Space Station}, submitted to Phys. Rev. Lett. 
\bibitem{AMS_ICRC_2015} Samuel C.C.Ting, {\em Latest Results from the Alpha Magnetic Spectrometer on the International Space Station}, 34th International Cosmic Ray Conference, July 30 to August 6 2015, The Hague, The Netherlands, to be published in the proceedings.
\bibitem{GEANT4} J. Allison et al., IEEE Trans. Nucl. Sci. 53 (2006) 270;
S. Agostinelli et al., Nucl. Instrum. Meth. A 506 (2003) 250.
\bibitem{L3_SMD} M. Acciarri et al., Nucl. Instr. Meth. A 351 (1994) 300
\bibitem{ATLAS_SCT} A. Ahmad et al., Nucl. Instr. Meth. A 578 (2007) 98
\bibitem{Fermi_LAT} W.B. Atwood et al., Astrophys. J. 697 (2009) 1071  
\bibitem{DAMPE} X. Wu et al, {\em The Silicon-Tungsten Tracker of the DAMPE Mission}, 34th International Cosmic Ray Conference, July 30 to August 6 2015, The Hague, The Netherlands, PoS (ICRC2015) 1192
\bibitem{Azzarello} Ph. Azzarello, {\em Tests And Production Of The AMS-02 Silicon Tracker Detectors}, PhD thesis No. 3530, Université de Genève 2004, \verb+http://dpnc.unige.ch/THESES/AZZARELLO_these.pdf+
\bibitem{Saouter} P. Saouter, {\em Nuclei Identification with the AMS-02 Silicon Tracker and Measurement of Nuclei Fluxes from Lithium to Neon}, PhD thesis No. 4706, Université de Genève 2014, \verb+http://dpnc.unige.ch/THESES/THESE_SAOUTER.pdf+
\bibitem{CREAM} E. S. Seo, Advances in Space Research 53 (2014) 1451.
\bibitem{CALET} P. Maestro, 23rd European Cosmic Ray Symposium, J.Phys. Conf. Ser. 409 (2013) 012026.
\bibitem{GAMMA-400} N.P. Topchiev et al., {\em GAMMA-400 gamma-ray observatory}, 34th International Cosmic Ray Conference, July 30 to August 6 2015, The Hague, The Netherlands, to be published in the proceedings
\bibitem{HERD} See { \verb+http://herd.ihep.ac.cn+}
\bibitem{PANGU} X. Wu et al, {\em PANGU: A High Resolution Gamma-Ray Space Telescope}, 34th International Cosmic Ray Conference, July 30 to August 6 2015, The Hague, The Netherlands, PoS (ICRC2015)964

\end{thebibliography}
\end{document}